\documentclass{ws-procs9x6-cpt19}
\begin{document}

\newcommand{\refeq}[1]{(\ref{#1})}
\def\etal {{\it et al.}}

\title{First Demonstration of Antimatter Quantum Interferometry}

\author{M.\ Giammarchi}

\address{Istituto Nazionale di Fisica Nucleare, Sezione di Milano,\\
Via Celoria 16, 20133, Italy}

\author{On behalf of the QUPLAS Collaboration}

\begin{abstract}
This paper describes the first experimental evidence of antimatter-wave
interference, a process at the heart of Quantum Physics and its interpretation.
For the case of ordinary matter particles, interference phenomena have been observed in a variety of cases, ranging from electrons up to complex molecules.
Here I present the first demonstration of single-positrons quantum interference.
\end{abstract}

\bodymatter

\section{Introduction}

The concept of wave-particle duality was introduced in 1923 by 
de Broglie\cite{DeBroglie1,DeBroglie2}: the Planck constant $h$ relates 
the momentum of a massive particle $p$ to a waveleght $\lambda_{dB}=h/p$.

Accordingly, diffraction and interference phenomena have been observed for 
a variety of particles, ranging from electrons\cite{Electrons1,Electrons2}, to
neutrons\cite{Neutrons1,Neutrons2}, and complex molecules\cite{Molecules1,Molecules2,Molecules3}. Gravitationally induced phase shifts were measured with neutrons in the famous Colella-Overhauser-Werner series of experiments\cite{COW1,COW2}.
 
The experimental study of single particle (one-at-a-time) double-slit interference was proposed by Feynman as a {\it gedanken} experiment and a decisive test that {\it has in it the heart of quantum mechanics}\cite{Feynman}.

The single electron interference experiment was made for the first time in 
1976\cite{MMP} and subsequently voted "the most beautiful experiment" by the readers of the Physics World magazine\cite{Crease}. A few years later, diffraction by positrons was observed for the first time\cite{Posdif}. A demonstration of single-particle interference for any antimatter particle was however still missing. 

In order to fill this gap, we have designed, realized and operated a Talbot-Lau
interferometer\cite{TL} suitable for anti-electrons (positrons) in the 5-18 keV energy range. This development is part of the QUPLAS (QUantum interferometry and
gravitation with Positrons and LASers) research program\cite{Theo1,Theo2,Exp1,Exp2}.

\begin{figure}
\begin{center}
\includegraphics[width=4in,height=2.5in]{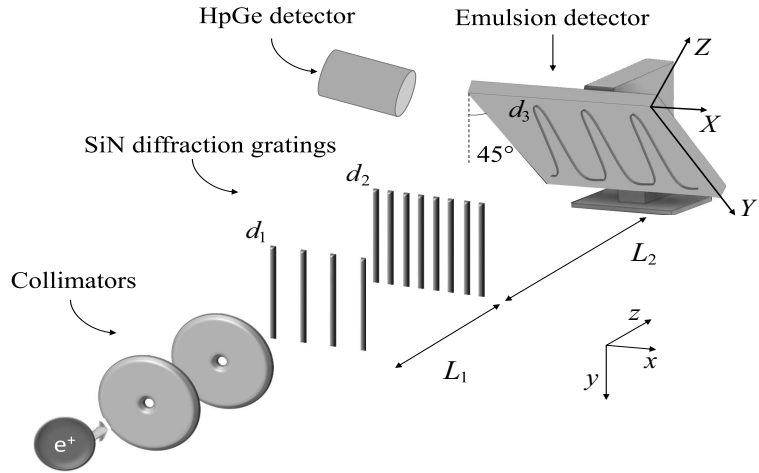}
\end{center}
\caption{Sketch of the setup: the beam, prepared by two 2-mm-wide collimators, reaches the gratings and the emulsion detector. In the setup, L$_1$=11.8 cm and L$_2$=57.6 cm. The emulsion is tilted by 45 degrees in order to better scan the longitudinal variation of fringes visibility. An HpGe detector is used to monitor the beam flux.}
\label{aba:fig1}
\end{figure}

\section{The setup}

The experiment is located at the variable energy positron beam facility of
L-NESS (Laboratory for Nanostructure Epitaxy and Spintronics on Silicon) of the Politecnico di Milano in Como (Italy). The positron beam has an intensity of $\sim 5\times 10^3$ $e^+/s$, an
energy between 5 and 18 keV (resolution better than 0.1\%) and an angular divergence of a few milliradians, producing a typical spot of about 2 $mm$. The beam is collimated and followed by the interferometer and the emulsion detector as shown in fig. \ref{aba:fig1}. The interferometer structure consists of two gratings in a period-magnifying Talbot-Lau configuration\cite{Theo2}. The two
gold-coated SiN gratings (11.8 $cm$ apart) have nominal periods of d$_1$=1.2 
and 
d$_2$=1 $\mu m$ respectively and a 50\% open fraction, producing a d$_3$=5.9 $\mu m$ period fringes at the location of the emulsion detector. 

The periodic spatial distribution is recorded by the emulsion detector that has a $\sim\mu m$ resolution. The emulsion has been tested at the QUPLAS energies in a dedicated "engineering" run\cite{Exp2} and the setup has been aligned and calibrated as described in [21].

The positron flux ($\sim10^3/s$) is produced by an incoherent (radioactive) source. Since the transit time of the particles in the interferometer is of  $\sim10^{-8}$ $s$, QUPLAS is clearly a one-particle-at-a-time experiment.

\begin{figure}
\begin{center}
\includegraphics[width=4in]{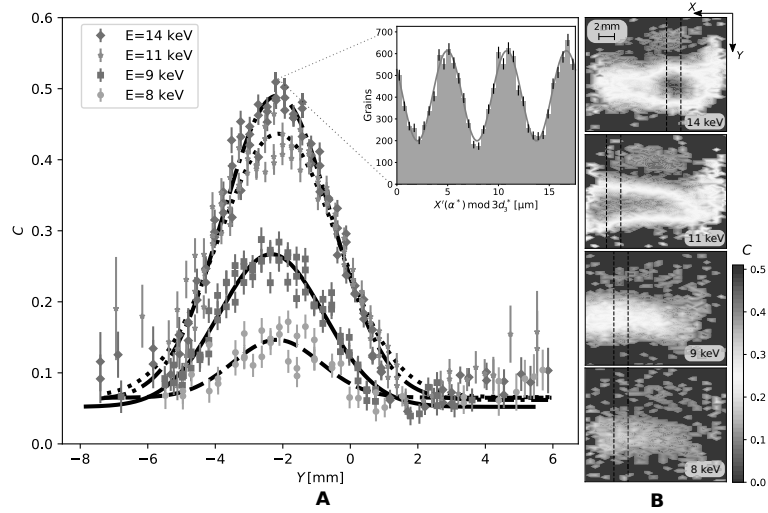}
\end{center}
\caption{A: Contrast C as a function of the logitudinal coordinate y for the different energies in the regions delimited by the dashed lines in B. The insert shows the actual periodicity for the case of best visibility. B: Contrast as a function of position on the emulsion for the different energies. See also fig. 1.}
\label{abb:fig2}
\end{figure}

\begin{figure}
\begin{center}
\includegraphics[width=4in,height=3.9in]{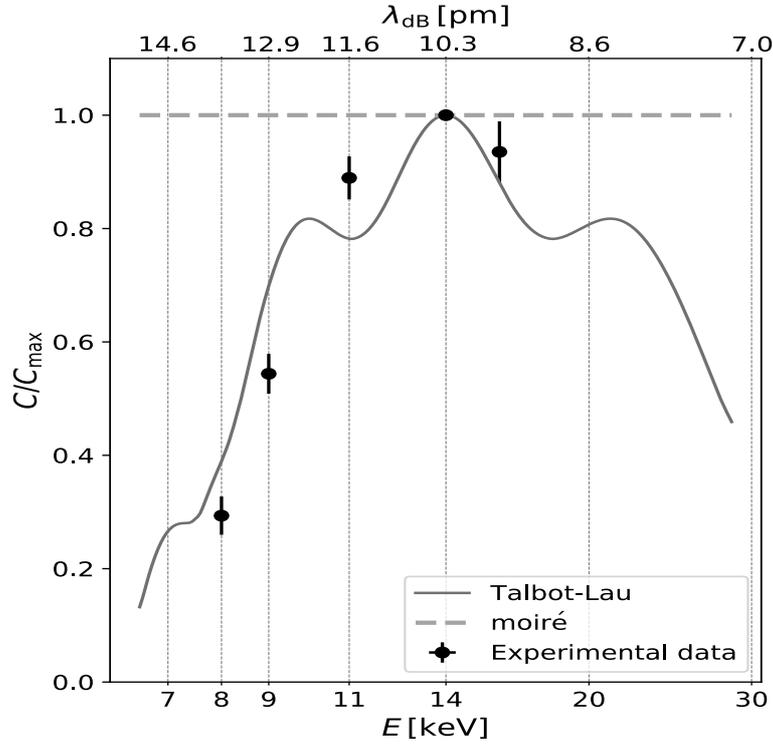}
\end{center}
\caption{The behaviour of the visibility as a function of energy proves that
the effect is of a quantum mechanical nature. The corrisponding classical projective (Moir\`e) would have been independent from the wavelength.}
\label{abc:fig3}
\end{figure}

\section{Results}

The configuration of the interferometer was meeting the resonant condition for the nominal 14 keV energy value, as\cite{Theo2} 

\begin{equation}
\frac{L_1}{L_2}=\frac{d_1}{d_2}-1
\label{fun:equ} 
\end{equation}

Exposures of the emulsions were made at 8, 9, 11, 14 and 16 keV positron energies; the results are presented in fig. \ref{abb:fig2}. For each energy, 
the emulsion was exposed for about 100 hours, then developed and the impact positions of positrons were digitized. The analysis strategy was to fit the fringes distribution as a function of both the period and the rotation angle between the interferometer and the emulsion detector (the remaining important alignment parameter). Periodical patterns as the one shown in the insert of fig. \ref{abb:fig2} has been obtained, with the expected periodicity of 5.9 $\mu m$. 

\section{Discussion}

After having shown periodical patterns as expected, in order to fully prove the quantum nature of the effect, the visibility was studied as a function of the energy (wavelength). As shown in fig. \ref{abc:fig3}, the fringes contrast as a function of energy disagrees with projective classical mechanics and is in agreement with the quantum mechanical model of the 
system\cite{Theo1,Theo2}.

This is the first demonstration of antimatter quantum interference.


\begin{thebibliography}{xx}

\bibitem{DeBroglie1}
{\it Waves and quanta,}
L. de Broglie, Nature 112 (1923) 140.

\bibitem{DeBroglie2}
{\it Recherches sul la theorie des quanta,}
L. de Broglie, Ann. Phys. 3 (1925) 22.

\bibitem{Electrons1}
{\it Reflection of electrons by a crystal of nickel,}
C.J. Davisson and L.H. Germer, Proc. Natl. Acad. Sci. USA 14 (1928) 317.

\bibitem{Electrons2}
{\it Diffraction of cathode rays by a thin film,} 
G.P. Thomson, A. Reid, Nature 119 (1927) 890.

\bibitem{Neutrons1}
{\it Test of a single crystal neutron interferometer,}
H. Rauch, W. Treimer and U. Bonse, Phys. Lett. A 47 (1974) 369. 

\bibitem{Neutrons2}
{\it Single- and double-slit diffraction of neutrons,}
A. Zeilinger, R. G\"ahler, C.G. Shull, W. Treimer and W. Mampe, Rev. Mod. Phys. 60 (1988) 1067.

\bibitem{Molecules1}
{\it Optics and interferometry with Na$_2$ molecules,}
M.S. Chapman, C.R. Ekstrom, T.D. Hammond, R.A. Rubenstein, J. Schmiedmayer,
S. Wehinger and D.E. Pritchard, Phys. Rev. Lett. 74 (1995) 4783.

\bibitem{Molecules2}
{\it Wave-particle duality of C$_{60}$ molecules,}
M. Arndt, O. Nairz, J. Vos-Andreae, C. Keller, G. van der Zouw and A. Zeilinger,
Nature 401 (1999) 680.

\bibitem{Molecules3} 
{\it Matter-wave interferometer for large molecules,}
B. Brezger, L. Hackerm\"uller, S. Uttenthaler, J. Petschinka, M. Arndt and A. Zeilinger, Phys. Rev. Lett. 88 (2002) 100404. 

\bibitem{COW1}
{\it Experimental Test of Gravitationally Induced Quantum Interference,}
A.W. Overhauser and R. Colella, Phys. Rev. Lett. 33 (1974) 1237.

\bibitem{COW2}
{\it Observation of Gravitationally Induced Quantum Interference,}
R. Colella, A.W. Overhauser and S.A. Werner, Phys. Rev. Lett. 34 (1975) 1472.

\bibitem{Feynman}
{\it Feyman Lectures on Physics Vol. 3}
Feynman, Leighton, Sands, 1965.

\bibitem{MMP}
{\it On the statistical aspect of electron interference phenomena,}
P.G. Merli, G.F. Missiroli and G. Pozzi, Am. J. Phys. 44 (1976) 306.

\bibitem{Crease}
{\it The most beautiful experiment,}
R.P. Crease, Phys. World 15 (2002) 19.

\bibitem{Posdif}
{\it Low energy positron diffraction from a Cu(111) surface,}
I.J. Rosenberg, A.H. Weiss and K.F. Canter, Phys. Rev. Lett. 44 (1980) 1139.

\bibitem{TL}
{\it Talbot-von Lau atom interferometry with cold slow potassium,}
J.F. Clauser and S. Li, Phys. Rev. A 49 (1994) R2213. 

\bibitem{Theo1}
{\it Matter-wave interferometry: towards antimatter interferometers,}
S. Sala, F. Castelli, M. Giammarchi, S. Siccardi and S. Olivares,
J. Phys. B 48 (2015) 195002.

\bibitem{Theo2}
{\it Asymmetric Talbot-Lau interferometry for intertial sensing,}
S. Sala, M. Giammarchi and S. Olivares,
Phys. Rev. A 94 (2016) 033625.

\bibitem{Exp1} 
{\it Detection of low energy antimatter with emulsions,}
S. Aghion, A. Ariga, T. Ariga, M. Bollani, E. Dei Cas, A. Ereditato, C. Evans, R. Ferragut, M. Giammarchi, C. Pistillo, M. Rome, S. Sala and P. Scampoli,
Journal of Instrumentation JINST 11 (2016) P06017.

\bibitem{Exp2}
{\it Nuclear emulsions for the detection of micrometric-scale fringe patterns:
an application to positron interferometry,}
S. Aghion, A. Ariga, M. Bollani A. Ereditato, R. Ferragut, M. Giammarchi, 
M. Lodari, C. Pistillo, S. Sala, P. Scampoli and M. Vladymyrov,
Journal of Instrumentation JINST 13 (2018) P05013.

\bibitem{SciAdv}
{\it First demonstration of antimatter wave interferometry,}
S. Sala, A. Ariga, A. Ereditato, R. Ferragut, M. Giammarchi, M. Leone, 
C. Pistillo and P. Scampoli,
Science Advances 5 eaav7610 (2019) doi: 10.1126/sciadv.aav7610.

\end{thebibliography}
\end{document}